# Reflexion Mössbauer analysis of the *in situ* oxidation products of hydroxycarbonate green rust

**S Naille, M Abdelmoula, D Louber, J-P Moulin and C Ruby**
Laboratoire de Chimie Physique et Microbiologie pour l'Environnement
LCPME UMR 7564 CNRS - Nancy-Université
405 rue de Vandœuvre, 54600 Villers-lès-Nancy, France

E-mail: sebastien.naille@lcpme.cnrs-nancy.fr

**Abstract.** The purpose of this study is to determine the nature of the oxidation products of $Fe^{II-III}$ hydroxycarbonate $Fe^{II}_4Fe^{III}_2(OH)_{12}CO_3 \cdot 3H_2O$ (green rust $GR(CO_3^{2-})$) by using the miniaturised Mössbauer spectrometer MIMOS II. Two Mössbauer measurements methods are used: method (i) with green rust pastes coated with glycerol and spread into Plexiglas sample holders, and method (ii) with green rust pastes in the same sample holders but introduced into a gas-tight cell with a beryllium window under a continuous nitrogen flow. Method (ii) allows us to follow the continuous deprotonation of $GR(CO_3^{2-})$ into the fully ferric deprotonated form $Fe^{III}_6O_4(OH)_8CO_3 \cdot 3H_2O$ by adding the correct amount of $H_2O_2$, without any further oxidation or degradation of the samples.

## 1. Introduction

The mixed-valence $Fe^{II}$-$Fe^{III}$ hydroxysalt compounds, named green rusts (GRs) due to their "bluish-green" colour [1], play an important role in environmental science due to their high redox flexibility and reactivity [2,3]. They were first identified as intermediate products in the corrosion process of iron-based materials [4] but are now known to reduce many anionic pollutants [5,6]. Considered for a long time as laboratory synthetic products, it is only ten years ago that GRs were discovered as minerals in hydromorphic gleysoil [7].

They belong to the layered double hydroxide (LDH) family with the general formula $[Fe^{II}_{(1-x)}Fe^{III}_x(OH)_2]^{x+} \cdot [(x/n)A^{n-} \cdot mH_2O]^{x-}$ (with $x = [Fe^{III}] / \{[Fe^{II}] + [Fe^{III}]\}$) which are made up of alternate brucite-type iron-hydroxide layers and interlayers consisting of water molecules and anions $A^{n-}$ such as $Cl^-$, $SO_4^{2-}$, $CO_3^{2-}$, ... Anions allow to establish the electroneutrality of these compounds as the iron-hydroxide layers are positively charged due to the presence of trivalent iron. Moreover, the crystal structure depends on the type of intercalated anions. When spherical (halogens) or planar (e.g. $CO_3^{2-}$) anions are intercalated, GRs crystallise in the trigonal system, space group $R\bar{3}m$, and are named green rusts one (GR1) [8,9] whereas green rusts two (GR2) crystallise in the hexagonal system, space group $P\bar{3}1m$, when tetragonal anions (e.g. $SO_4^{2-}$) are intercalated [10].

GRs are involved in the formation of iron oxides and (oxy)hydroxides in the natural environment. Since these discoveries, several works have been done on their oxidation processes [11,12]. The oxidation products depend on several factors such as oxidation rate and kinetic, nature of the GR, pH, temperature or $Fe^{II}$ concentration. The most common products formed by GRs oxidation are lepidocrocite, goethite or ferrihydrite. However, recent studies have shown that $Fe^{II}$-$Fe^{III}$ hydroxycarbonate $Fe^{II}_4Fe^{III}_2(OH)_{12}CO_3 \cdot 3H_2O$ (stoichiometric green rust $GR(CO_3^{2-})$, $x = 0.33$) leads to the formation of $Fe^{III}$ oxyhydroxycarbonate $Fe^{III}_6O_4(OH)_8CO_3 \cdot 3H_2O$ ($GR(CO_3^{2-})^*$, $x = 1$) by fast and/or violent oxidation [12,13]. The proposed mechanism of the $GR(CO_3^{2-})$ oxidation mode with $H_2O_2$ is a solid-state reaction due to the *in situ* progressive deprotonation of the hydroxyl sheets within the layered structure. The ferric molar ratio $x$ is continuously rising with the increasing amount of $H_2O_2$ for the $Fe^{II}$-$Fe^{III}$ oxyhydroxycarbonate intermediates $Fe^{II}_{6(1-x)}Fe^{III}_{6x}O_{12}H_{2(7-3x)}CO_3 \cdot 3H_2O$ ($GR(CO_3^{2-})^*$, $x \in ]0.33\text{-}1[$) [13]. Nevertheless, except for the fully ferric form ($x = 1$), $Fe^{II}$-$Fe^{III}$ hydroxycarbonate green rusts are air sensitive due to the divalent iron content. The aim of this study is to show how it is possible to monitor at room temperature the oxidation products of $GR(CO_3^{2-})$ with $H_2O_2$ by means of the Mössbauer spectrometer MIMOS II [14] without any further oxidation.

## 2. Experimental

$GR(CO_3^{2-})$ was synthesised by coprecipitation method. First, $Fe^{II\text{-}III}$ solution ($[Fe^{II}] + [Fe^{III}] = 0.4$ M and $x = 0.33$) was prepared by dissolving weighted amounts of $FeSO_4 \cdot 7H_2O$ (Aldrich, 99+%) and $Fe_2(SO_4)_3 \cdot 5H_2O$ (Aldrich, 97%) with 100 mL of demineralised water in an Erlenmeyer flask. Moreover, an additional amount of $NaH_2PO_4$ (Aldrich, 99%) was dissolved in the mixture in order to stabilise the $GR(CO_3^{2-})$ by the mechanism of phosphate adsorption [15]. Following this, a 100 mL basic solution of 0.8 M NaOH (Aldrich, 97+%) containing dissolved $Na_2CO_3$ (Sigma-Aldrich, 99.5+%) with $[CO_3^{2-}] = 0.466$ M, was rapidly transferred to the first mixture and a "bluish-green" precipitate appeared immediately. $CO_3^{2-}$ ions were added in excess compared to $SO_4^{2-}$ ions to prevent the formation of hydroxysulphate green rust $GR(SO_4^{2-})$. Finally, the erlenmeyer flask was filled with demineralised water and sealed off to avoid any air oxidation.

$GR(CO_3^{2-})^*$, with different $x$ values, were obtained by violent oxidation of $GR(CO_3^{2-})$ with a $H_2O_2$ solution (Sigma-Aldrich, 30wt.%). Calculated volumes of $H_2O_2$ were dropped under vigorous stirring to reach four different compositions ($x = 0.5, 0.67, 0.83$ and $1$) according to the following equation:

$$Fe^{II}_4Fe^{III}_2(OH)_{12}CO_3 \cdot 3H_2O + (3x-1)H_2O_2 \rightarrow Fe^{II}_{6(1-x)}Fe^{III}_{6x}O_{12}H_{2(7-3x)}CO_3 \cdot 3H_2O + 2(3x-1)H_2O$$

The increasing oxidation of divalent iron can be followed by the precipitates changing colour, which gradually transforms from "bluish-green" ($x = 0.33$) to "brownish-orange" ($x = 1$).

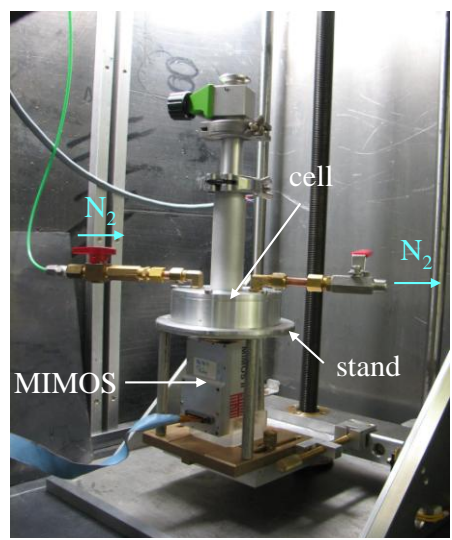

**Figure 1.** MIMOS II against the gas-tight cell supported by a three-legged stand.

The room temperature (RT) $^{57}$Fe Mössbauer spectra were recorded in backscattered measurement geometry in the constant acceleration mode using the miniaturised Mössbauer spectrometer MIMOS II, which was designed and fabricated at the University of Mainz (Germany) [14]. The source used for experiments was $^{57}$Co (50 mCi) embedded in a Rh matrix and the velocity scale was calibrated with the magnetic sextet of a high purity iron foil as the reference absorber. The spectra were fitted to either Lorentzian shape lines or pseudo-Voigt profile analysis with the Recoil© program [16] and the quality of the fits were controlled by the usual $\chi^2$ test. Two methods of sample preparation were used to avoid any air oxidation, but in both cases the precipitates were first filtered and the wet green rusts pastes were spread on the Mylar window of a machined Plexiglas cell as sample holder. Method (i) consisted to coat every paste with glycerol and the sample holder was placed on the spectrometer head right above the four Si-PIN-diode detectors. For method (ii), the sample holders were incorporated into a home-made gas-tight cell with a beryllium window and the data recording was performed under a continuous nitrogen flow (figure 1).

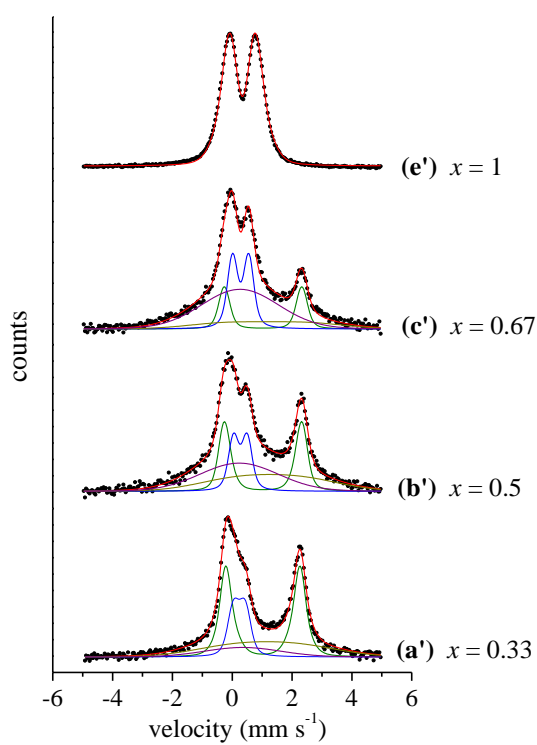

**Figure 2.** Backscattered Mössbauer spectra of $GR(CO_3^{2-})$ at $x = 0.33$ and $GR(CO_3^{2-})^*$ oxidised at different values of the ferric molar fraction $x$. The spectra are recorded with method (i) at room temperature.

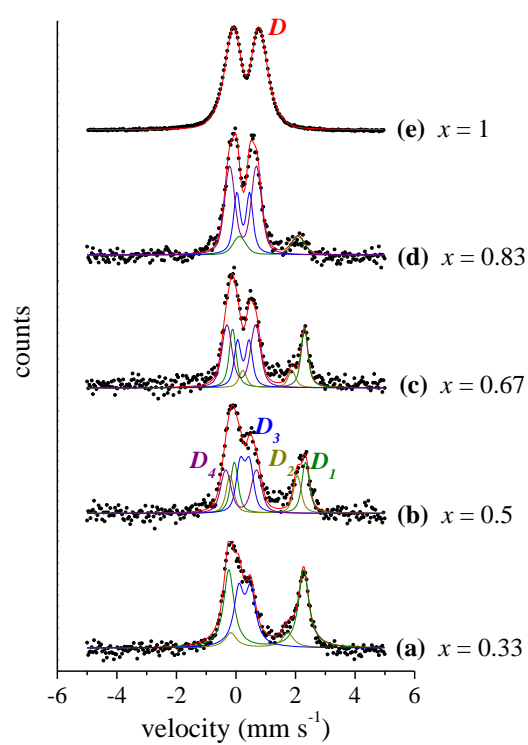

**Figure 3.** Backscattered Mössbauer spectra of $GR(CO_3^{2-})$ at $x = 0.33$ and $GR(CO_3^{2-})^*$ oxidised at different values of the ferric molar fraction $x$. The spectra are recorded with method (ii) at room temperature.

### 3. Results and discussion

The Mössbauer spectra of $GR(CO_3^{2-})$ and its oxidation products using method (i), are shown in figure 2. This sampling method, consisting of coating the pastes to avoid air oxidation, is often used for the X-ray diffraction (XRD) characterisation of LDHs [12,15]. However, the Mössbauer spectra exhibit a huge broadening, leading to a very large line widths as a result of the Mössbauer fitting procedure. Although the $Fe^{II}$ components intensities decrease with the increasing added amount of $H_2O_2$, these spectra are not exploitable. This is due to glycerol which interacts chemically with LDHs by acting as a swelling agent [17]. Since glycerol is simply spread out on the GRs pastes, the reaction between glycerol and GRs takes time and thus explain why there is no evidence in the XRD patterns which are generally recorded in one hour. In the case of Mössbauer spectrometry, and especially in backscattering geometry, a spectrum can be recorded within four days. The reaction between GRs and glycerol can then occur and may lead to a delamination process which falsifies the Mössbauer spectra results.

In order to tackle this problem, another method, method (ii), has been used to characterise the oxidation products of GR($CO_3^{2-}$). A gas-tight cell with a beryllium window has been designed (figure 1) at the laboratory to record spectra at room temperature under a continuous nitrogen flow (figure 3). The spectrum of GR($CO_3^{2-}$) (figure 3(a)) is fitted with two ferrous doublets ($D_1$ and $D_2$) and one ferric doublet ($D_3$); the hyperfine parameters are summarised in table 1. The use of two doublets for ferrous cations and only one for ferric cations is consistent with the environment surrounding these cations. Indeed, $Fe^{II}$ cations can be close to either water molecules ($Fe^{II}(H_2O)$) or $CO_3^{2-}$ anions ($Fe^{II}(CO_3^{2-})$) whereas $Fe^{III}$ cations environment cannot be easily distinguished [18]. The experimental abundances observed for the three different sites by Mössbauer spectroscopy (table 1) are in good agreement with the theoretical ones which are 1/2, 1/6 and 1/3 for $Fe^{II}(CO_3^{2-})$, $Fe^{II}(H_2O)$ and $Fe^{III}$ respectively. The centre shifts (CS) values of the different doublets confirm that Fe–O bonds in the brucite-type layers, where each cation lies at the centre of an octahedron with $OH^-$ ions at its six vertices, are highly ionic due to the inductive effect. The non-zero values of the quadrupole splittings ($\Delta$) are characteristic of high-spin iron species in a distorted octahedral configuration as explained in [19].

**Table 1.** Mössbauer hyperfine parameters of the spectra presented in figure 3: the different components used in the Mössbauer fitting procedure and their relative area (R.A.), the ferric molar fraction ($x = [Fe^{III}] / [Fe_{total}]$) either measured by Mössbauer spectroscopy or targeted from the added quantity of $H_2O_2$, the centre shift (CS) (Lorentzian shape lines, figure 2(a)) or the centre shift ($<CS>$) at the maximum of the Gaussian distribution (Voigt profiles, figures 2(b-e)) with respect to metallic α-Fe at room temperature and the quadrupole splitting ($\Delta$) or the quadrupole splitting at the maximum of the Gaussian distribution ($<\Delta>$).

| Fig. 3 | Component | R.A. (%) | $x$ (measured) | $x$ (targeted) | CS or $<CS>$ (mm s$^{-1}$) | $\Delta$ or $<\Delta>$ (mm s$^{-1}$) |
|---|---|---|---|---|---|---|
| (a) | $D_1$ ($Fe^{II}$) | 54 | 0.35 | 0.33 | 1.13 | 2.51 |
|  | $D_2$ ($Fe^{II}$) | 11 |  |  | 0.90 | 1.91 |
|  | $D_3$ ($Fe^{III}$) | 35 |  |  | 0.40 | 0.40 |
| (b) | $D_1$ ($Fe^{II}$) | 26 | 0.54 | 0.5 | 1.25 | 2.39 |
|  | $D_2$ ($Fe^{II}$) | 18 |  |  | 1.04 | 2.26 |
|  | $D_3$ ($Fe^{III}$) | 26 |  |  | 0.41 | 0.31 |
|  | $D_4$ ($Fe^{III}$) | 28 |  |  | 0.29 | 1.03 |
| (c) | $D_1$ ($Fe^{II}$) | 28 | 0.63 | 0.67 | 1.20 | 2.42 |
|  | $D_2$ ($Fe^{II}$) | 9 |  |  | 1.16 | 1.65 |
|  | $D_3$ ($Fe^{III}$) | 22 |  |  | 0.36 | 0.39 |
|  | $D_4$ ($Fe^{III}$) | 41 |  |  | 0.29 | 0.96 |
| (d) | $D_1$ ($Fe^{II}$) | 15 | 0.85 | 0.83 | 1.21 | 1.97 |
|  | $D_3$ ($Fe^{III}$) | 28 |  |  | 0.35 | 0.43 |
|  | $D_4$ ($Fe^{III}$) | 57 |  |  | 0.35 | 0.90 |
| (e) | $D$ ($Fe^{III}$) | 100 | 1 | 1 | 0.45 | 0.98 |

The formation of $Fe^{II}$-$Fe^{III}$ oxyhydroxycarbonate green rusts GR($CO_3^{2-}$)* as intermediates with $x \in ]0.33-1[$ by violent oxidation of GR($CO_3^{2-}$) is confirmed by the MIMOS spectra (figures 3(b-d)) and the hyperfine parameters (table 1). Voigt profile analyses are used to fit every spectrum due to a global broadening which has already been observed by using transmission Mössbauer spectrometry [18]. A second ferric doublet ($D_4$) with a high quadrupole splitting appears and is in good agreement with the progressive deprotonation of the brucite-type layers, which was observed by XPS study [13], with a distortion of the octahedral sites that results from the mixture of $OH^-$ and $O_2^-$ ions at the octahedra vertices. Furthermore, the relative area of $D_4$ compared to $D_3$ is continuously increasing and ferrous doublets, $D_1$ and $D_2$, are gradually vanishing. The Mössbauer spectrometry at room temperature makes it thus possible to validate the *in situ* deprotonation process of the GR($CO_3^{2-}$) hydroxide layers with $H_2O_2$ since there is a good agreement between the targeted $x$ values and those measured (table 1). The fully ferric sample GR(CO$_3^{2-}$)* ($x = 1$) is fitted with a broad Fe$^{III}$ paramagnetic doublet (*D*), covering every Fe$^{III}$ sites that can exist in the cationic layers (figure 3(e)). This shows that GR(CO$_3^{2-}$) can be fully oxidised in accordance with a solid-state reaction without any dissolution processes taking place.

**4. Conclusion**

Depending on the preparation method, significant differences were observed on the MIMOS spectra for GR(CO$_3^{2-}$) and its *in situ* oxidation products with H$_2$O$_2$. These oxidation products are Fe$^{II}$-Fe$^{III}$ oxyhydroxycarbonate intermediates and include the fully ferric deprotonated salt at the final stage of the oxidation process. Method (i) leads to a significant broadening of the spectra that could be attributed to a chemical interaction between GRs and glycerol. Such a broadening decreases significantly when the experiments are done under a controlled atmosphere. The ferric molar ratio $x = $ [Fe$^{III}$] / {[Fe$^{II}$] + [Fe$^{III}$]}, determined by using the miniaturised MIMOS(II) spectrometer, are then very close to the expected $x$ values that take into account the amount of added H$_2$O$_2$.

**Acknowledgments**
The authors would like to acknowledge Nadia Sabri (McGill University, Montreal, Quebec, Canada) for her constructive comments, which improved the quality of the manuscript.